\documentclass[12pt]{article}
\usepackage{amssymb}
\begin{document}



\def\a{\alpha}
\def\b{\beta}
\def\d{\delta}
\def\e{\epsilon}
\def\g{\gamma}
\def\h{\mathfrak{h}}
\def\k{\kappa}
\def\l{\lambda}
\def\o{\omega}
\def\p{\wp}
\def\r{\rho}
\def\t{\theta}
\def\s{\sigma}
\def\z{\zeta}
\def\x{\xi}
 \def\A{{\cal{A}}}
 \def\B{{\cal{B}}}
 \def\C{{\cal{C}}}
 \def\D{{\cal{D}}}
\def\G{\Gamma}
\def\K{{\cal{K}}}
\def\O{\Omega}
\def\R{\bar{R}}
\def\T{{\cal{T}}}
\def\L{\Lambda}
\def\f{E_{\tau,\eta}(sl_2)}
\def\E{E_{\tau,\eta}(sl_n)}
\def\Zb{\mathbb{Z}}
\def\Cb{\mathbb{C}}

\def\R{\overline{R}}

\def\beq{\begin{equation}}
\def\eeq{\end{equation}}
\def\bea{\begin{eqnarray}}
\def\eea{\end{eqnarray}}
\def\ba{\begin{array}}
\def\ea{\end{array}}
\def\no{\nonumber}
\def\le{\langle}
\def\re{\rangle}
\def\lt{\left}
\def\rt{\right}

\newtheorem{Theorem}{Theorem}
\newtheorem{Definition}{Definition}
\newtheorem{Proposition}{Proposition}
\newtheorem{Lemma}{Lemma}
\newtheorem{Corollary}{Corollary}
\newcommand{\proof}[1]{{\bf Proof. }
        #1\begin{flushright}$\Box$\end{flushright}}

\baselineskip=20pt

\newfont{\elevenmib}{cmmib10 scaled\magstep1}
\newcommand{\preprint}{
   \begin{flushleft}
   \end{flushleft}\vspace{-1.3cm}
   \begin{flushright}\normalsize
     {\tt hep-th/yymmnnn} \\ December 2006
   \end{flushright}}
\newcommand{\Title}[1]{{\baselineskip=26pt
   \begin{center} \Large \bf #1 \\ \ \\ \end{center}}}
\newcommand{\Author}{\begin{center}
   \large \bf
Wen-Li Yang ${}^{a,b}$,
  Yao-Zhong Zhang ${}^{b,c}$ and  Xin Liu ${}^b$\end{center}}
\newcommand{\Address}{\begin{center}

${}^a$ Institute of Modern Physics, Northwest University,
       Xian 710069, P.R. China\\
${}^b$ Department of Mathematics, University of Queensland, Brisbane,
       QLD 4072, Australia\\
${}^c$ Condensed Matter Theory Laboratory, RIKEN, Wako, Saitama
351-0198, Japan\\
E-mail: \,wenli@maths.uq.edu.au,\,\,
yzz@maths.uq.edu.au,\,\,xinliu@maths.uq.edu.au

   \end{center}}
\newcommand{\Accepted}[1]{\begin{center}
   {\large \sf #1}\\ \vspace{1mm}{\small \sf Accepted for Publication}
   \end{center}}

\preprint
\thispagestyle{empty}
\bigskip\bigskip\bigskip

\Title{Free field realization of current superalgebra $gl(m|n)_k$}
\Author

\Address
\vspace{1cm}

\begin{abstract}
We construct the free field representation of the affine currents,
energy-momentum tensor and screening currents of
the first kind of the current superalgebra $gl(m|n)_k$  uniformly for
$m=n$ and $m\neq n$.  The energy-momentum tensor is
given by a linear combination of two Sugawara tensors associated
with the two independent quadratic Casimir elements of $gl(m|n)$.

\vspace{1truecm} \noindent {\it PACS:} 11.25.Hf; 02.20.Tw

\end{abstract}
\newpage
\section{Introduction}
\label{intro} \setcounter{equation}{0}

Current superalgebras or affine superalgebras have emerged in a wide
range of physical areas ranging from
high energy physics to condensed matter physics. In high energy theoretical
physics, sigma models with supermanifold target spaces naturally appear
in the quantization of superstring theory in the AdS-type backgrounds.
It was argued in \cite{Bers99} that even without a WZNW term
the sigma model on $PSL(n|n)$ supergroup is already conformally
invariant. A WZNW term with any integer coefficient may then be added
whenever necessary, without violating the conformal invariance.
In \cite{Ber99},  the $PSU(1,1|2)$ sigma model was used to quantize superstring
theory on the AdS$_3\times S^3$ background with Ramond-Ramond (RR) flux.
In condensed matter physics, the supersymmetric treatment of quenched disorders
leads to current superalgebras of zero superdimension. It is believed
that critical behaviours of certain disordered systems
such as the integer quantum Hall transition are described by
sigma models or their WZNW generalizations
based on supergroups of zero superdimension
\cite{Ber95,Mud96,Maa97,Bas00,Gur00}.

As can be seen from the work \cite{Ber00}, in most above-mentioned applications
one expects to work with sigma models on some kind of coset supermanifolds
\cite{Met98,Roi00} with a WZNW term, and thus models of interest
are more complicated than WZNW models on non-coset supergroups. However,
even for such (non-coset) supergroup WZNW models, little has been known
in general \cite{Sch06}, largely due to the technical difficulties in
dealing with atypical and indecomposable representations which are common
features for most superalgebras.

The Wakimoto free field realization \cite{Wak86,Fra97} has been
proved to be a powerful method in the study of CFTs such as WZNW
models. Motivated by the above-mentioned applications, in this
paper  we construct free field representations of the current
superalgebra $gl(m|n)_k$ associated with the $GL(m|n)$ WZNW model
for $m=n$ and $m\neq n$ in a unified way.

Free field realization of the $gl(m|n)$ currents, {\it in
principle\/}, can be obtained by a general method outlined in
\cite{Fei90,Bou90,Ito90,Fre94,Ras98}, where differential
realizations of the corresponding finite dimensional Lie (super)
algebras play a key role. However, {\it in practice}, such
constructions for an {\it explicit} expression of the currents
become very complicated for higher-rank algebras \cite{Ito90,
Ras98, Din03, Zha05}. In this paper, we find a way to overcome the
complication. In our approach, the construction of the
differential operator realization becomes much simpler (c.f.
\cite{Ras98, Din03}). We demonstrated this by working out the
differential realization of $gl(4|4)$ in \cite{Yan06}. Here we
provide the complete results of $gl(m|n)$ for any $m$ and $n$.

This paper is organized as follows. In section 2, we briefly
review the definitions of finite-dimensional superalgebra
$gl(m|n)$ and the associated current algebra, which also services
as introducing our notation and some basic ingredients. In section
3, we construct explicitly the differential operator realization
of $gl(m|n)$ in the standard basis. In section 4 and section 5, we
construct the free field representation of the affine currents
associated with $gl(m|n)$ at a generic level $k$, and the
corresponding energy-momentum tensor. We moreover construct, in
section 6, the screening currents of the first kind. Section 7 is
for conclusions.


\section{Notation and prelimilaries}
\label{CUR} \setcounter{equation}{0}

Let us first fix our notation for the underlying non-affine
superalgebra $gl(m|n)$  which is non-semisimple for both $m=n$ and
$m\neq n$ \cite{Kac77,Fra96}.

$gl(m|n)$ is $\Zb_2$ graded and is
generated by the elements $\{E_{i,j}|\,i,j=1,\ldots,m+n\}$ which satisfy
the following (anti)commutation relations:
\bea [E_{i,j}, E_{k,l}]=
\d_{j\,k}E_{i,l}-(-1)^{([i]+[j])([k]+[l])}\d_{i\,l}E_{k,j}.
\label{Def-1}\eea
Here and throughout, we adopt the convention:
$[a,b]=ab-(-1)^{[a][b]}ba$.  The $\Zb_2$ grading of the generators
is $[E_{i,j}]=[i]+[j]$ with $[1]=\ldots=[m]=0,\,[m+1]=\ldots=[m+n]=1$.
$E_{i,j},\; 1\leq i\neq j\leq m+n$, are raising/lowering generators.
For a unified treatment of the $m=n$ and $m\neq n$ cases, we have chosen
$E_{i,i},\; i=1,\cdots, m+n$, to be the elements of the
Cartan subalgebra (CSA) of $gl(m|n)$.

Let us remark that other bases of the CSA widely used by most physicists
do not seem suitable for the unified treatment because the CSA elements
for $m=n$ and $m\neq n$ in those bases are different. This is seen as follows.
Let
$$ I=\sum_{i=1}^{m+n}E_{ii},~~~~~~ J=\sum_{i=1}^{m+n}(-1)^{[i]}E_{ii}.$$
In the fundamental representation of $gl(m|n)$, $I$ is the $(m+n)\times
(m+n)$ identity matrix and $J$ is the diagonal matrix
$J=diag(1,\cdots,1,-1,\cdots,-1)$. For $m\neq n$ the usual choice
of the $gl(m|n)$ CSA elements is $\{I, H_i=(-1)^{[i]}E_{i,i}-(-1)^{[i+1]}
E_{i+1,i+1}, \; i=1,\cdots,m+n-1\}$. However, this choice is inappropriate
for $m=n$ because in this case $I$ and $H_i,\;i=1,\cdots,2n-1$
become dependent:
$$ H_n=\frac{1}{n}\lt\{I-\sum_{i=1}^{n-1}[iH_i+(n-i)H_{n+i}]\rt\}.$$
That is, for $m=n$ one can not simutaneously choose, say $H_n$ and $I$ as part
of the $gl(n|n)$ CSA elements, in contrast to the $m\neq n$ case.
One popular choice of $gl(n|n)$ CSA elements is
$\{I, J, H_i,\;1\leq i\neq n\leq 2n-1\}$. On the other hand, for $m\neq n$,
$J$ is a linear combination of $I$ and $H_i,\;1\leq i\leq m+n-1$.
Note also that for $m=n$, $I\in sl(n|n)$ and both $sl(n|n)$ and $gl(n|n)$
are non-semisimple; for $m\neq n$, $I$ is in $gl(m|n)$ but not in $sl(m|n)$.

With our above choice of generators, it is easy to check that the usual
quadratic Casimir element of $gl(m|n)$ is
\bea
C_1=\sum_{i,j=1}^{m+n}\,(-1)^{[j]}E_{i,j}E_{j,i}.\label{Casim-1}
\eea
Since $gl(m|n)$ is non-semisimple, there exists another independent quadratic
Casimir element
\bea
C_2=\sum_{i,j=1}^{m+n}\,E_{i,i}E_{j,j}=\lt(\sum_{i=1}^{m+n}E_{i,i}\rt)^2.
\label{Casim-2}
\eea These two Casimir elements are useful in the
following for the construction of the correct energy-momentum tensor.


For any $m$ and $n$, $gl(m|n)$ has a non-degenerate and invariant metric
or bilinear form given by \cite{Kac77,Fra96} \bea
\lt(E_{i,j},\,E_{k,l}\rt)=str\lt(e_{i,j}e_{k,l}\rt).
\label{Inner-product}\eea
Here $e_{i,j}$, which is the $(m+n)\times(m+n)$ matrix with entry
$1$ at the $i$-th row and $j$-th column and zero elsewhere, is the
fundamental or defining representation of $E_{i,j}$;
$str$ denotes the supertrace, i.e., $str(a)=\sum_{i}(-1)^{[i]}\,a_{i\,i}$.

The $gl(m|n)$ current algebra is generated by the currents
$E_{i,j}(z)$  associated with the generators $E_{i,j}$ of
$gl(m|n)$. The current algebra at a general level $k$ obeys the
following OPEs \cite{Fra97}, \bea E_{i,j}(z)\,E_{l,m}(w)=
k\frac{\lt(E_{i,j},E_{l,m}\rt)}{(z-w)^2}+\frac{1}{(z-w)}
\lt(\d_{j\,l}\,E_{i,m}(w)-(-1)^{([i]+[j])([l]+[m])}\d_{i\,m}\,
E_{l,j}(w)\rt) .\label{OPE}\eea

\section{Differential operator realization of $gl(m|n)$}
 \label{DIR} \setcounter{equation}{0}

As mentioned in the introduction, practically it would be very
involved (if not impossible) to obtain the {\it explicit} free
field realization of  higher-rank algebras such as $gl(m|n)$ for a
larger value of $m+n$ by the general method outlined in
\cite{Fei90,Bou90,Ito90,Fre94,Ras98}. We have found a way to
overcome the complication. In our approach, the construction of
the differential operator realization becomes much simpler.

Let us introduce $\frac{1}{2}((m(m-1)+n(n-1))$ bosonic coordinates
$\{x_{i,j},\,x_{m+k,m+l}|\,1\leq i<j\leq m,\,1\leq k<l\leq n\}$
with the $\Zb_2$-grading: $[x_{i,j}]=0$, and $m\times n$ fermionic
coordinates $\{\t_{i,m+j}|\,1\leq i\leq m,\, 1\leq j\leq n\}$ with
the $\Zb_2$-grading: $[\t_{i,m+j}]=1$. These coordinates satisfy
the following (anti)commutation relations: \bea
&&[x_{i,j},x_{k,l}]=0,\,\,
[\partial_{x_{i,j}},x_{k,l}]=\d_{ik}\d_{jl},\no\\
&&[\t_{i,m+j},\t_{k,m+l}]=0,\,\,
[\partial_{\t_{i,m+j}},\t_{k,m+l}]=\d_{ik}\d_{jl},\no\eea and the
other (anti)commutation relations are vanishing. Let $\langle\L|$
be the lowest weight vector in the associated representation of
$gl(m|n)$, satisfying the following conditions: \bea
&&\langle\L|E_{j+1,j}=0,\quad 1\leq j\leq
m+n-1,\label{Lowestweight-1}\\
&&\langle\L|E_{i,i}=\l_i\,\langle\L|,\quad 1\leq i\leq
m+n.\label{Lowestweight-2}\eea An arbitrary vector in this
representation space is parametrized by $\langle\L|$ and the
coordinates ($x$ and $\t$)  as \bea
\langle\L,x,\t|=\langle\L|G_{+}(x,\t),\label{States}\eea where
$G_{+}(x,\t)$ is given by (c.f. \cite{Ras98}) \bea G_{+}(x,\t)=
(G_{m+n-1,m+n})\ldots (G_{j,m+n}\ldots G_{j,j+1})\ldots \,
(G_{1,m+n}\ldots G_{1,2}).\eea Here, for $i<j$, $G_{i,j}$ are
given by \bea
G_{i,j}=\lt\{\begin{array}{ll}e^{x_{i,j}E_{i,j}},&{\rm
if}\,\, [E_{i,j}]=0,\\
e^{\t_{i,j}E_{i,j}},&{\rm if}\,\, [E_{i,j}]=1.\end{array}\rt.\eea
One can define a differential operator realization $\rho^{(d)}$ of
the generators of $gl(m|n)$ by \bea
\rho^{(d)}(g)\,\langle\L,x,\t|\equiv \langle\L,x,\t|\,g,\qquad
\forall g\in gl(m|n).\label{definition} \eea Here $\rho^{(d)}(g)$
is a differential operator of the bosonic and ferminic coordinates
associated with the generator $g$, which can be obtained from the
defining relation (\ref{definition}). Moreover, the defining
relation also assure the differential operator realization is
actually a representation of $gl(m|n)$. Therefore it is sufficient
to obtain the associated differential operators which are related
to the simple roots, and the others can be constructed through the
simple ones by (anti)commutation relations
(\ref{Non-simple-1})-(\ref{Non-simple-2}) (see below). By using
the relation (\ref{definition}) and the Baker-Campbell-Hausdorff
formula, after some algebraic manipulations, we obtain the following
differential operator representation of the simple
generators in the standard (distinguished) basis \cite{Fra96}:
\bea \rho^{(d)}(E_{j,j+1})&=&\sum_{k\leq
j-1}x_{k,j}\,\partial_{x_{k,j+1}}+
\partial_{x_{j,j+1}},\,\,1\leq j\leq m-1, \label{Diff-1}\\
\rho^{(d)}(E_{m,m+1})&=&\sum_{k\leq
m-1}x_{k,m}\,\partial_{\t_{k,m+1}}
+\partial_{\t_{m,m+1}}, \\
\rho^{(d)}(E_{m+j,m+1+j})&=&\sum_{k\leq
m}\t_{k,m+j}\,\partial_{\t_{k,m+1+j}}+ \sum_{k\leq
j-1}x_{m+k,m+j}\,\partial_{x_{m+k,m+1+j}}+
\partial_{x_{m+j,m+1+j}},\no\\
&&\qquad 1\leq j\leq n-1,\\
\rho^{(d)}(E_{j,j})&=&\sum_{k\leq j-1}x_{k,j}\,\partial_{x_{k,j}}-
\sum_{j+1\leq k\leq m}x_{j,k}\,\partial_{x_{j,k}} -\sum_{k\leq
n}\t_{j,m+k}\,\partial_{\t_{j,m+k}}+\l_j,\no\\
&&\qquad 1\leq j\leq m-1,\\
\rho^{(d)}(E_{m,m})&=&\sum_{k\leq
m-1}x_{k,m}\,\partial_{x_{k,m}}-\sum_{k\leq
n}\t_{m,m+k}\,\partial_{\t_{m,m+k}}+\l_m,\\
\rho^{(d)}(E_{m+j,m+j})&=&\sum_{k\leq
m}\t_{k,m+j}\,\partial_{\t_{k,m+j}}+
\sum_{k\leq j-1}x_{m+k,m+j}\,\partial_{x_{m+k,m+j}}\no\\
&&\qquad\qquad- \sum_{j+1\leq k\leq
n}x_{m+j,m+k}\,\partial_{x_{m+j,m+k}}+\l_{m+j},\,\,1\leq j\leq n,\\
\rho^{(d)}(E_{j+1,j})&=&\sum_{k\leq j-1}
x_{k,j+1}\,\partial_{x_{k,j}}-\sum_{j+2\leq k\leq m} x_{j,k}\,
\partial_{x_{j+1,k}}-\sum_{k\leq n}\t_{j,m+k}\,\partial_{\t_{j+1,m+k}} \no\\
&&\quad -x_{j,j+1}\lt(\sum_{j+1\leq k\leq
m}x_{j,k}\,\partial_{x_{j,k}}+\sum_{k\leq n}
\t_{j,m+k}\,\partial_{\t_{j,m+k}}\rt)\no\\
&&\quad +x_{j,j+1}\lt(\sum_{j+2\leq k\leq
m}x_{j+1,k}\,\partial_{x_{j+1,k}}+\sum_{k\leq
n}\t_{j+1,m+k}\,\partial_{\t_{j+1,m+k}}\rt)\no\\
&&\quad+x_{j,j+1} \lt(\l_j-\l_{j+1}\rt),\,\,1\leq j\leq m-1,\\
\rho^{(d)}(E_{m+1,m})&=&\sum_{k\leq
m-1}\t_{k,m+1}\,\partial_{x_{k,m}}+\sum_{2\leq
k\leq n}\t_{m,m+k}\partial_{x_{m+1,m+k}}\no\\
&&\quad-\t_{m,m+1}\lt(\sum_{2\leq k\leq
n}\lt(\t_{m,m+k}\,\partial_{\t_{m,m+k}}
+x_{m+1,m+k}\,\partial_{x_{m+1,m+k}}\rt)\rt)\no\\
&&\quad +\t_{m,m+1}\lt(\l_{m}+\l_{m+1}\rt),\\
\rho^{(d)}(E_{m+1+j,m+j})&=&\sum_{k\leq
m}\t_{k,m+1+j}\,\partial_{\t_{k,m+j}} +\sum_{k\leq
j-1}x_{m+k,m+1+j}\,\partial_{x_{m+k,m+j}}\no\\
&&\quad -\sum_{j+2\leq k\leq n}
x_{m+j,m+k}\,\partial_{x_{m+1+j,m+k}}\no\\
&&\quad -x_{m+j,m+1+j}\sum_{j+1\leq
k\leq n}x_{m+j,m+k}\,\partial_{x_{m+j,m+k}}\no\\
&&\quad +x_{m+j,m+1+j}\sum_{j+2\leq k\leq
n}x_{m+1+j,m+k}\,\partial_{x_{m+1+j,m+k}}\no\\
&&\quad +x_{m+j,m+1+j}\lt(\l_{m+j}-\l_{m+1+j}\rt),\,\,1\leq j\leq
n-1.\label{Diff-2}\eea
The generators associated with the
non-simple roots can be constructed through the simple ones
by the (anti)commutation relations, \bea
\rho^{(d)}(E_{i,j})&=&[\rho^{(d)}(E_{i,k}),\rho^{(d)}(\,E_{k,j})],\,\,
1\leq i<k<j\leq m+n\,\,{\rm and}\,2\leq j-i,\label{Non-simple-1}\\
\rho^{(d)}(E_{j,i})&=&[\rho^{(d)}(E_{j,k}),\,\rho^{(d)}(E_{k,i})],\,\,
1\leq i<k<j\leq m+n \,\,{\rm and}\,2\leq j-i.\label{Non-simple-2}
\eea
A direct computation shows that the differential
realization (\ref{Diff-1})-(\ref{Non-simple-2}) of $gl(m|n)$
satisfies the commutation relations (\ref{Def-1}). Alternatively, one
may check that (\ref{Diff-1})-(\ref{Diff-2})
satisfy the (anti)commutation relations corresponding to
the simple roots together with the Serre relations \cite{Fra96}.

\section{Free field realization of $gl(m|n)_k$}
 \label{FFR} \setcounter{equation}{0}

With the help of the differential realization obtained in the last
section we can construct the free field representation
of the $gl(m|n)$ current algebra in
terms of $\frac{1}{2}(m(m-1)+n(n-1))$ bosonic $\b$-$\g$ pairs
$\{(\b_{i,j},\,\g_{i,j}),\, 1\leq i<j\leq m;
(\bar{\b}_{i,j}\bar{\g}_{i,j}),\, 1\leq i<j\leq n\}$,
$m\times n$ fermionic $b$-$c$ pairs $\{(\psi^{\dagger}_{i,j},\,\psi_{i,j}),\,
1\leq i\leq m,\, 1\leq j\leq n\}$ and $m+n$ free scalar fields
$\phi_i$, $i=1,\ldots,m+n$. These free fields obey the
following OPEs:\bea
&&\hspace{-0.8truecm}\b_{i,j}(z)\,\g_{k,l}(w)=-\g_{k,l}(z)\,\b_{i,j}(w)=
\frac{\d_{ik}\d_{jl}}{(z-w)},\,\,1\leq i<j\leq m,\,\,1\leq
k<l\leq m,\label{OPE-F-1}\\
&&\hspace{-0.8truecm}\bar{\b}_{i,j}(z)\,\bar{\g}_{k,l}(w)=-\bar{\g}_{k,l}(z)\,
\bar{\b}_{i,j}(w)= \frac{\d_{ik}\d_{jl}}{(z-w)},\,\,1\leq i<j\leq
n,\,\,1\leq k<l\leq n,\\
&&\hspace{-0.8truecm}\psi_{i,j}(z)\psi_{k,l}^{\dagger}(w)
=\psi_{k,l}^{\dagger}(z)
\psi_{i,j}(w)=\frac{\d_{ik}\d_{jl}}{(z-w)},\,\,\,\,\,\,1\leq
i,k\leq m,\,\,1\leq j,l\leq n,\\
&&\hspace{-0.8truecm}\phi_i(z)\phi_j(w)=(E_{i,i}, E_{j,j})\ln(z-w)=
   (-1)^{[i]}\,\d_{ij}\,
\ln(z-w),\,\,\,\,\,1\leq i,j\leq m+n,\label{OPE-F-2}\eea
and the other OPEs are trivial.

The free field realization of the $gl(m|n)$ current algebra
(\ref{OPE}) is obtained by the substitution in the
differential realization (\ref{Diff-1})-(\ref{Diff-2}) of $gl(m|n)$,
\bea
&&x_{i,j}\longrightarrow \g_{i,j}(z),\quad \partial_{x_{i,j}}
\longrightarrow \b_{i,j}(z),\quad 1\leq i<j\leq m,\no\\
&&x_{m+i,m+j}\longrightarrow \bar{\g}_{i,j}(z),\quad
\partial_{x_{m+i,m+j}} \longrightarrow \bar{\b}_{i,j}(z),
\quad 1\leq i<j\leq n,\no\\
&&\t_{i,m+j}\longrightarrow \psi^{\dagger}_{i,j}(z),\quad
\partial_{\t_{i,m+j}} \longrightarrow \psi_{i,j}(z),\quad 1\leq i\leq
m\,{\rm and}\,1\leq j\leq n, \no\\
&&\l_j\longrightarrow \sqrt{k+m-n}\partial\phi_j(z)-
\frac{(-1)^{[j]}(1+\a)}{2\sqrt{k+m-n}}\sum_{l=1}^{m+n}\phi_l(z),\,\,
1\leq j\leq m+n,\no\eea  with
$\a=1+\frac{2k}{m-n}-\frac{2\sqrt{k(k+m-n)}}{m-n}$,
followed by the addition of anomalous terms linear in
$\partial \psi^{\dagger}(z)$, $\partial\g(z)$ and
$\partial\bar{\g}(z)$ in the expressions of the currents. It is
remarked that for $m=n$, $\a$ is $\a=\lim_{m\rightarrow
n}(1+\frac{2k}{m-n}-\frac{2\sqrt{k(k+m-n)}}{m-n})=0$. Here we
present the realization of the currents associated with the simple roots,
\bea E_{j,j+1}(z)&=&\sum_{l\leq
j-1}\g_{l,j}(z)\b_{l,j+1}(z)+\b_{j,j+1}(z),\,\,1\leq j\leq
m-1,\label{Fre-currents-1}\\
E_{m,m+1}(z)&=&\sum_{l\leq m-1}\g_{l,m}(z)\psi_{l,1}(z)+\psi_{m,1}(z),\\
E_{m+j,m+j+1}(z)&=&\sum_{l\leq
m}\psi^{\dagger}_{l,j}(z)\psi_{l,j+1}(z)+\sum_{l\leq
j-1}\bar{\g}_{l,j}(z)\bar{\b}_{l,j+1}(z)+\bar{\b}_{j,j+1}(z),\no\\
&&\qquad 1\leq j\leq n-1,\\
E_{j,j}(z)&=&\sum_{l\leq j-1}\g_{l,j}(z)\b_{l,j}(z)- \sum_{j+1\leq
l\leq m}\g_{j,l}(z)\b_{j,l}(z) -\sum_{l\leq
n}\psi^{\dagger}_{j,l}(z)\psi_{j,l}(z)\no\\
&&\quad+\sqrt{k+m-n}\partial\phi_j(z)-\frac{1+\a}{2\sqrt{k+m-n}}
\sum_{l=1}^{m+n}\partial\phi_l(z),\no\\
&&\qquad 1\leq j\leq m,\label{CSA-currents1}\\
E_{m+j,m+j}(z)&=&\sum_{l\leq
m}\psi^{\dagger}_{l,j}(z)\psi_{l,j}(z)+ \sum_{l\leq
j-1}\bar{\g}_{l,j}(z)\bar{\b}_{l,j}(z)- \sum_{j+1\leq l\leq
n}\bar{\g}_{j,l}(z)\bar{\b}_{j,l}(z)\no\\
&&\quad+\sqrt{k+m-n}\partial\phi_{m+j}(z)+\frac{1+\a}{2\sqrt{k+m-n}}
\sum_{l=1}^{m+n}\partial\phi_l(z),\no\\
&&\qquad 1\leq j\leq n, \label{CSA-currents2}\\
E_{j+1,j}(z)&=&\sum_{l\leq j-1}
\g_{l,j+1}(z)\b_{l,j}(z)-\sum_{j+2\leq l\leq m} \g_{j,l}(z)
\b_{j+1,l}(z)-\sum_{l\leq n}\psi^{\dagger}_{j,l}(z)\psi_{j+1,l}(z) \no\\
&&\quad -\g_{j,j+1}(z)\lt(\sum_{j+1\leq l\leq
m}\g_{j,l}(z)\b_{j,l}(z)+\sum_{l\leq n}
\psi^{\dagger}_{j,l}(z)\psi_{j,l}(z)\rt)\no\\
&&\quad +\g_{j,j+1}(z)\lt(\sum_{j+2\leq l\leq
m}\g_{j+1,l}(z)\b_{j+1,l}(z)+\sum_{l\leq
n}\psi^{\dagger}_{j+1,l}(z)\psi_{j+1,l}(z)\rt)\no\\
&&\quad+\sqrt{k+m-n}\g_{j,j+1}(z)
\lt(\partial\phi_j(z)-\partial\phi_{j+1}(z)\rt)
+(k+j-1)\partial\g_{j,j+1}(z),\no\\
&&\quad\qquad\,\,1\leq j\leq m-1,\\
E_{m+1,m}(z)&=&\sum_{l\leq
m-1}\psi^{\dagger}_{l,1}(z)\b_{l,m}(z)+\sum_{2\leq
l\leq n}\psi^{\dagger}_{m,l}(z)\bar{\b}_{1,l}(z)\no\\
&&\quad-\psi^{\dagger}_{m,1}(z)\lt(\sum_{2\leq l\leq
n}\lt(\psi^{\dagger}_{m,l}(z)\,\psi_{m,l}(z)
+\bar{\g}_{1,l}(z)\bar{\b}_{1,l}(z)\rt)\rt)\no\\
&&\quad +\sqrt{k+m-n}\psi^{\dagger}_{m,1}(z)
\lt(\partial\phi_{m}(z)+\partial\phi_{m+1}(z)\rt)\no\\
&&\quad+(k+m-1)\partial\psi^{\dagger}_{m,1}(z),\\
E_{m+j+1,m+j}(z)&=&\sum_{l\leq
m}\psi^{\dagger}_{l,j+1}(z)\psi_{l,j}(z) +\sum_{l\leq
j-1}\bar{\g}_{l,j+1}(z)\bar{\b}_{l,j}(z)-\sum_{j+2\leq
l\leq n} \bar{\g}_{j,l}(z)\bar{\b}_{j+1,l}(z)\no\\
&&\quad -\bar{\g}_{j,j+1}(z)\lt(\sum_{j+1\leq l\leq
n}\bar{\g}_{j,l}(z)\bar{\b}_{j,l}(z)-\sum_{j+2\leq l\leq
n}\bar{\g}_{j+1,l}(z)\bar{\b}_{j+1,l}(z)\rt)\no\\
&&\quad +\sqrt{k+m-n}\bar{\g}_{j,j+1}
\lt(\partial\phi_{m+j}(z)-\partial\phi_{m+j+1}(z)\rt)\no\\
&&\quad-(k+m+1-j)\partial\bar{\g}_{j,j+1}(z), \,\,1\leq j\leq
n-1.\label{Fre-currents-2}\eea
Here and throughout normal ordering of free
fields is implied whenever necessary. The free field realization of
currents associated with the non-simple roots can be obtained from
the OPEs of the simple ones, similar to
(\ref{Non-simple-1})-(\ref{Non-simple-2}). It is straightforward
to check that the above free field realization of the currents
satisfy the OPEs of the $gl(m|n)$ current algebra.
Moreover, for the case $n=0$  our results reduce to those in
\cite{Bou90}, giving the free field realization of the $gl(m)$ current algebra.

Some remarks are in order. We have obtained the free field realization of
$gl(m|n)$ current algebra uniformly for any $m$ and $n$ for the CSA basis
we have chosen. It is easy to make simple basis transforms of the CSA
to get expressions for the more familar CSA bases. This is seen as follows.
Introduce new free scalar fields through linear combinations of the
original free scalar fields $\phi_i(z)$,
\bea
\phi_I(z)&=&\sum_{i=1}^{m+n}\phi_i(z),~~~~~\phi_J(z)=\sum_{i=1}^{m+n}
   (-1)^{[i]}\phi_i(z),\no\\
\phi_{H_i}(z)&=&(-1)^{[i]}\phi_i(z)-(-1)^{[i+1]}\phi_{i+1}(z).
\eea
In terms of the new scalar fields, the currents associated with $I,J$ and
$H_i$ take the form, as can easily be seen from (\ref{CSA-currents1}) and
(\ref{CSA-currents2}),
\bea
I(z)&=&\sum_{i=1}^{m+n}E_{i,i}(z)=\sqrt{k}\partial\phi_I(z),\no\\
J(z)&=&\sum_{i=1}^{m+n}(-1)^{[i]}E_{i,i}(z)=\sqrt{k+m-n}\partial\phi_J(z)
  -\frac{(m+n)(1+\alpha)} {2\sqrt{k+m-n}}\partial\phi_I(z),\no\\
& &-\sum_{j=1}^n\sum_{l\leq n}\psi^\dagger_{jl}(z)\psi_{jl}(z)
   -\sum_{j=1}^n\sum_{l\leq m}\psi^\dagger_{lj}(z)\psi_{lj}(z),\no\\
H_i(z)&=&(-1)^{[i]}E_{i,i}(z)-(-1)^{[i+1]}E_{i+1,i+1}(z)\no\\
&=&\tilde{H}_i(z)+\sqrt{k+m-n}\partial\phi_{H_i}(z),~~~1\leq i\leq m+n-1,
\eea
where $\tilde{H}_i(z)$ are functions of the $\beta$-$\gamma$ and
$b$-$c$ pairs only. Now for $m=n$, replacing the $2n$ original free scalar
fields $\phi_i(z)$ by $\{\phi_{H_i}(z),\;1\leq i\neq n\leq 2n-1, \phi_I(z),
\phi_J(z)\}$ and moreover using the relation,
$$\phi_{H_n}(z)=\frac{1}{n}\lt[\phi_I(z)-\sum_{i=1}^{n-1}\lt(i\phi_{H_i}(z)
   +(n-i)\phi_{H_{n+i}}(z)\rt)\rt]$$
to eliminate $\phi_{H_n}(z)$, then we obtain the $gl(n|n)$ currents
$\{E_{i,j}(z),\;1\leq i\neq j\leq 2n-1; I(z), J(z), H_l(z),\;1\leq l
 \neq n\leq 2n-1\}$ in the new basis in terms of the new free scalar fields
defined above together with the original $\beta$-$\gamma$ and $b$-$c$ pairs.
Similarly for $m\neq n$, we replace the $m+n$ original free scalar fields
$\phi_i(z)$ by  $\{\phi_{H_i}(z),\;1\leq i\leq m+n-1, \phi_I(z)\}$ to
obtain the $gl(m|n)$ currents $\{E_{i,j}(z),\;1\leq i\neq j\leq 2n-1; I(z),
H_l(z),\;1\leq l\leq m+n-1\}$ in the new basis.

Note that for $m=n$, $\phi_J(z)$ only appears in $J(z)$. Thus the free
field realization of $sl(n|n)$ current algebra may be obtained from
that of the $gl(n|n)$ current algebra  by simply dropping $J(z)$.
The free field realization of $psl(n|n)=sl(n|n)/I$ current algebra
is obtained by setting $\phi_I(z)=0$ and thus $I(z)=0$ in the realization
of the $sl(n|n)$ current algebra. For $m\neq n$, since
$\phi_I(z)$ only appears in $I(z)$ in the new basis,
one may obtain the free field realization of the $sl(m|n)$ current algebra
by simply dropping $I(z)$ in the realization of the $gl(m|n)$ current algebra.


\section{Energy-momentum tensor}
\label{EMT} \setcounter{equation}{0}

In this setion we construct the free field realization of the Sugawara
energy-momentum tensor associated with the $gl(m|n)$ current algebra.
After a tedious calculation, we find that the Sugawara tensor corresponding
to the quadratic Casimir $C_1$ is given by \bea
T_1(z)&=&\frac{1}{2(k+m-n)}\sum_{i,j=1}^{m+n}(-1)^{[j]}:
E_{i,j}(z)E_{j,i}(z):\no\\
&=&\frac{1}{2}\sum_{l=1}^{m+n}(-1)^{[l]}\partial\phi_l(z)\partial\phi_l(z)\no\\
&&\quad-\frac{1}{2\sqrt{k+m-n}}\partial^2
\lt(\sum_{i=1}^m(m-n-2i+1)\phi_i(z)-\sum_{j=1}^n(m+n-2j+1)\phi_{m+j}(z)\rt)\no\\
&&\quad+\sum_{i<j}^m\partial\g_{i,j}(z)\b_{i,j}(z)
+\sum_{i<j}^n\partial\bar{\g}_{i,j}(z)\bar{\b}_{i,j}(z)\no\\
&&\quad+\sum_{i=1}^m\sum_{j=1}^n\partial\psi^{\dagger}_{i,j}(z)\psi_{i,j}(z)
-\frac{1}{2(k+m-n)}\partial\phi_I(z)\,\partial\phi_I(z).\eea
On the other hand,  the Sugawara tensor corresponding to the
quadratic Casimir $C_2$ is  \bea
T_2(z)&=&\frac{1}{2(k+m-n)}\sum_{i,j=1}^{m+n}:E_{i,i}(z)E_{j,j}(z):\no\\
&=&\frac{k}{2(k+m-n)}\partial\phi_I(z)\,\partial\phi_I(z).\eea In
order that all currents $E_{i,j}(z)$ are primary fields with
conformal dimensional one, we define the energy-momentum tensor
$T(z)$ as follow: \bea T(z)&=&T_1(z)+\frac{1}{k}T_2(z)\no\\
&=&\frac{1}{2}\sum_{l=1}^{m+n}(-1)^{[l]}\partial\phi_l(z)\partial\phi_l(z)\no\\
&&\quad-\frac{1}{2\sqrt{k+m-n}}\partial^2
\lt(\sum_{i=1}^m(m-n-2i+1)\phi_i(z)-\sum_{j=1}^n(m+n-2j+1)\phi_{m+j}(z)\rt)\no\\
&&\quad+\sum_{i<j}^m\partial\g_{i,j}(z)\b_{i,j}(z)
+\sum_{i<j}^n\partial\bar{\g}_{i,j}(z)\bar{\b}_{i,j}(z)
+\sum_{i=1}^m\sum_{j=1}^n\partial\psi^{\dagger}_{i,j}(z)\psi_{i,j}(z).
\label{Energy-Momentum}\eea
It is straightforward to check that
$T(z)$ satisfy the following OPE, \bea
T(z)T(w)=\frac{c/2}{(z-w)^4}+\frac{2T(w)}{(z-w)^2}+\frac{\partial
T(w)}{(z-w)},\eea
where the  central charge $c=0$ for $m=n$, and
\bea
c=
\frac{k\lt((m-n)^2-1\rt)}{k+m-n}+1
\label{Center-charge}\eea
for $m\neq n$. Moreover, we find that with regard to
the energy-momentum tensor $T(z)$ defined by
(\ref{Energy-Momentum}) all currents $E_{i,j}(z)$ are indeed
primary fields with conformal dimensional one, namely, \bea
T(z)E_{i,j}(w)=\frac{E_{i,j}(w)}{(z-w)^2}+\frac{\partial
E_{i,j}(w)}{(z-w)},\,\,1\leq i,j\leq m+n.\eea Therefore, $T(z)$ is
the very energy-momentum tensor of the $gl(m|n)$ current algebra.



\section{Screening currents}
\label{SC}  \setcounter{equation}{0}

Important objects in applying the free field realization to the
computation of correlation functions  of the associated CFT are
screening currents. A screening current is a primary field with
conformal dimension one and has the property that the singular
part of its OPE with the affine currents is a total derivative.
These properties ensure that integrated screening currents
(screening charges) may be inserted into correlators while the
conformal or affine Ward identities remain intact. This in turn
makes them very useful in the computation of correlation functions
\cite{Dos84,Ber90}.

Free field realization of the $gl(m|n)$ screening
currents of the first kind may be constructed from certain differential
operators $\{s_{i,j}|1\leq i<j\leq n+m\}$ \cite{Bou90,Ras98} which we
define as given from the relation
\bea s_{i,j}\,
\langle\L,x,\t|\equiv\langle\L|\,E_{i,j}\,G_{+}(x,\t),\qquad {\rm
for}\, 1\leq i,j\leq m+n.\label{Def-2}\eea
The above-defined operators $s_{i,j}$ give a differential
operator realization of a subalgebra of $gl(m|n)$.
Again it is sufficient to construct $s_{j,j+1}$ related to the
simple generators $E_{j,j+1}$, $1\leq j\leq m+n-1$ of $gl(m|n)$.
Let us denote those differential operators by $s_{j}$. Using
(\ref{Def-2}) and the Baker-Campbell-Hausdorff formula,
after some algebraic manipulations, we obtain the following explicit
expressions of $s_j$:
\bea s_j&=& \sum_{j+2\leq l\leq
m}x_{j+1,l}\partial_{x_{j,l}}+\sum_{l\leq
n}\theta_{j+1,m+l}\partial_{\theta_{j,m+l}}+\partial_{x_{j,j+1}},\quad
1\leq j\leq m-1,\\
s_m&=&\sum_{2\leq l\leq n}
x_{m+1,m+l}\partial_{\theta_{m,m+l}}+\partial_{\theta_{m,m+1}},\\
s_{m+j}&=&\sum_{j+2\leq l\leq
n}x_{m+j+1,m+l}\partial_{x_{m+j,m+l}}+\partial_{x_{m+j,m+j+1}},\quad
1\leq j\leq n-1.\eea
{}From (anti)communication relations similar to
(\ref{Non-simple-1}), one may obtain the differential operators $s_{i,j}$
associated with the non-simple generators of $gl(m|n)$.
Following the procedure similar to \cite{Bou90,Ras98},
we find the free field realization of the screening currents $S_j$
corresponding to the differential operators $s_j$,
\bea S_j(z)&=&\lt(\sum_{j+2\leq l\leq
m}\g_{j+1,l}(z)\b_{j,l}(z)
+\sum_{l=1}^n\psi^{\dagger}_{j+1,l}(z)\psi_{j,l}(z)+\b_{j,j+1}(z)\rt)
\tilde{s}_j(z),\no\\
&&\qquad\qquad\,\, 1\leq j\leq m-1,\label{Screening-1}\\
S_m(z)&=&\lt(\sum_{2\leq
l\leq n}\bar{\g}_{1,l}(z)\psi_{m,l}(z)+\psi_{m,1}(z)\rt)\tilde{s}_m(z),\\
S_{m+j}(z)&=&\lt(\sum_{j+2\leq l\leq
n}\bar{\g}_{j+1,l}(z)\bar{\b}_{j,l}(z)+\bar{\b}_{j,j+1}(z)\rt)
\tilde{s}_{m+j}(z) ,\,\,1\leq j\leq n-1,\eea where \bea
&&\tilde{s}_j(z)=e^{-\frac{1}{\sqrt{k+m-n}}(\phi_j(z)-\phi_{j+1}(z))},\quad
1\leq j\leq m-1,\\
&&\tilde{s}_m(z)=e^{-\frac{1}{\sqrt{k+m-n}}(\phi_{m}(z)+\phi_{m+1}(z))},\\
&&\tilde{s}_{m+j}(z)=e^{\frac{1}{\sqrt{k+m-n}}(\phi_{m+j}(z)-\phi_{m+j+1}(z))},
\quad 1\leq j\leq n-1. \label{Screening-2}\eea
The OPEs of the screening currents with the energy-momentum tensor
and the $gl(m|n)$ currents (\ref{Fre-currents-1})-(\ref{Fre-currents-2}) are
\bea &&
T(z)S_j(w)=\frac{S_j(w)}{(z-w)^2}+\frac{\partial S_j(w)}{(z-w)}
=\partial_w\lt\{\frac{S_j(w)}{(z-w)}\rt\},\,\,1\leq j\leq m+n-1,\\
&&E_{i+1,i}(z)S_j(w)=(-1)^{[i]+[i+1]}\d_{ij}\,
\partial_{w}\lt\{\frac{k \,\tilde{s}_j(w)}{(z-w)}\rt\},\,\,
1\leq i,j\leq m+n-1,\\
&&E_{i,i+1}(z)S_j(w)=0,\,\, 1\leq i,j\leq m+n-1,\\
&&E_{i,i}(z)S_j(w)=0,\,\, 1\leq i\leq m+n,\,\, 1\leq j\leq
m+n-1.\eea
The screening currents obtained this way are screening currents
of the first kind \cite{Ber86}. Moreover, $S_m(z)$ is
fermionic and the others are bosonic.


\section{Discussions}
\label{Con} \setcounter{equation}{0}

We have studied the $gl(m|n)$ current algebra at general level
$k$. We have constructed its Wakimoto free field realization
(\ref{Fre-currents-1})-(\ref{Fre-currents-2}) for $m=n$ and
$m\neq n$ in a unified way, and the
corresponding  energy-momentum tensor (\ref{Energy-Momentum})
which is a linear combination of two Sugawara tensors associated
with two quadratic Casimir elements of $gl(m|n)$.
We have also found $m+n-1$ screening currents,
(\ref{Screening-1})-(\ref{Screening-2}), of the first kind.
Our results reduce to those in \cite{Bou90} for $n=0$ (i.e. in the
bosonic case), and recover those in \cite{Yan06} for $m=n=4$, thus
providing a complete proof of the results in that paper.

To fully take the advantage of the free field approach in applications
mentioned in the introduction, explicit construction of primary fields
in terms of free fields is needed. It is well-known that there exist
two types of representations for the underlying finite dimensional
superalgebra $gl(m|n)$: typical and atypical representations.
Atypical representations, which are often indecomposable,
have no counterparts in the bosonic
algebra setting and the understanding of such representations is
still very much incomplete. Although the construction of the
primary fields associated with typical representations are similar
to the bosonic algebra cases, it is a highly non-trivial task to
construct the primary fields associated with atypical
representations \cite{Zha05}.

\section*{Acknowledgements}
We are happy to thank Akira Furusaki for helpful discussions. YZZ would
like to thank the Condensed Matter Theory Laboratory,
especially Akira Furusaki, of the RIKEN Discovery Research Institute
for warm hospitality and financial support.
This work has been supported by the Australian Research Council.
XL gratefully acknowledges the support of the IPRS and UQGSS
scholarships of the University of Queensland.

\vspace{1.00truecm}

\noindent{{\large \it Note added}}: \, We became aware that free
field realization of $sl(m|n)$  current algebra was investigated
previously in \cite{Bar91} (for $gl(n|n)$ case see also
\cite{Isi94}). There, the $sl(m|n)$ currents were expressed in
terms of  $sl(m)$ and $sl(n)$ currents with different levels and
some $b$-$c$ pairs. As part of the results of our paper, we give
the {\em explicit} expressions of $gl(m|n)$ currents in terms of
free fields, by using a different method.



\begin{thebibliography}{99}
\bibitem{Bers99} M. Bershadsky, S. Zhukov and A. Vaintrob, {\it
Nucl. Phys.\/} {\bf B 559} (1999), 205.
\bibitem{Ber99} N. Berkovits, C. Vafa and E. Witten, {\it JHEP\/},
{\bf 9903} (1999), 018.
\bibitem{Ber95} D. Bernard, {\tt hep-th/9509137}.
\bibitem{Mud96} C. Mudry, C. Chamon and X.\,-G. Wen, {\it Nucl.
Phys.\/} {\bf B 466} (1996), 383.
\bibitem{Maa97} Z. Massarani and D. Serban, {\it Nucl. Phys.\/}
{\bf B 489} (1997), 603.
\bibitem{Bas00} Z.\,S. Bassi and A. LeClair, {\it Nucl. Phys.\/}
{\bf B 578} (2000), 577.
\bibitem{Gur00} S. Guruswamy, A. LeClair and A.\,W.\,W. Ludwig,
{\it Nucl. Phys.\/} {\bf B 583} (2000), 475.
\bibitem{Ber00} N. Berkovits, M. Bershadsky, T. Hauer, S. Zhukov
and B. Zwiebach, {\it Nucl. Phys.\/} {\bf B 567} (2000), 61.
\bibitem{Met98} R.\,R. Metsaev and A.\,A. Tseytlin, {\it Nucl.
Phys.\/} {\bf 533} (1998), 109.
\bibitem{Roi00} R. Roiban and W. Siegel, {\it JHEP\/}, {\bf 0011}
(2000), 024.
\bibitem{Sch06} V. Schomerus and H. Saleur, {\it Nucl. Phys.\/}
{\bf B 734} (2006), 221.
\bibitem{Wak86} M. Wakimoto, {\it Commun. Math. Phys.\/} {\bf 104}
(1986), 605.
\bibitem{Fra97} P. Di Francesco, P. Mathieu and D. Senehal, {\it
Conformal Field Theory\/}, Springer Press, Berlin, 1997.
\bibitem{Fei90} B. Feigin and E. Frenkel, {\it Commun. Math.
Phys.\/} {\bf 128} (1990), 161.
\bibitem{Bou90} P. Bouwknegt, J. McCarthy and K. Pilch, {\it Prog.
Phys. Suppl.\/} {\bf  102} (1990), 67.
\bibitem{Ito90} K. Ito, {\it Phys. Lett.\/} {\bf B 252} (1990), 69.
\bibitem{Fre94} E. Frenkel, {\tt hep-th/9408109}.
\bibitem{Ras98} J. Rasmussen, {\it Nucl. Phys.\/} {\bf B 510} (1998), 688.
\bibitem{Din03} X.\,-M. Ding, M. Gould and Y.\,-Z. Zhang, {\it
Phys. Lett.\/} {\bf A 318} (2003), 354.
\bibitem{Zha05} Y.\,-Z. Zhang, X. Liu and W.\,-L. Yang, {\it Nucl.
Phys.\/} {\bf B 704} (2005), 510.
\bibitem{Yan06} W.\,-L. Yang, Y.\,-Z. Zhang and X. Liu, {\it Phys. Lett. }
{\bf B 641} (2006), 329.
\bibitem{Kac77}V.\,G. Kac, {\it Adv. Math.\/} {\bf 26} (1977), 8.
\bibitem{Fra96} L. Frappat, P. Sorba and A. Sciarrino, Dictionary
on Lie algebras and superalgebras, Academic Press, New York, 2000.



\bibitem{Dos84} VI.\,S. Dotsenko and V.\,A. Fateev, {\it Nucl.
Phys.\/} {\bf B 240} (1984), 312.
\bibitem{Ber90} D. Bernard and G. Felder, {\it Commun. Math.
Phys.\/} {\bf 127} (1990), 145.
\bibitem{Ber86} M. Bershadsky and H. Ooguri, {\it Commun. Math.
Phys.\/} {\bf 126} (1986), 49.

\bibitem{Bar91} I. Bars, {\it Phys. Lett.\/} {\bf B 255} (1991),
353.
\bibitem{Isi94} J.\,M. Isidro and A.\,V. Ramallo, {\it Nucl. Phys.
\/} {\bf B 414} (1994), 715.

\end{thebibliography}
\end{document}